\documentclass[aps,prl,preprint,groupaddress]{revtex4}
\usepackage[dvipdfmx]{graphicx}
\usepackage{bm}
\usepackage{pifont}
\usepackage{amssymb, amsmath}
\usepackage{dcolumn}
\usepackage{color}

\begin{document}
\title{Bulk quantum Hall effect of spin-valley-coupled Dirac fermions in a polar antiferromagnet BaMnSb$_2$
}
\author{H. Sakai$^{1,2}$}\email[Corresponding author: ]{sakai@phys.sci.osaka-u.ac.jp}
\author{H. Fujimura$^1$}
\author{S, Sakuragi$^3$}
\author{M. Ochi$^1$}
\author{R. Kurihara$^3$}
\author{A. Miyake$^3$}
\author{M. Tokunaga$^3$}
\author{T. Kojima$^4$}
\author{D. Hashizume$^5$}
\author{T. Muro$^6$}
\author{K. Kuroda$^3$}
\author{T. Kondo$^3$}
\author{T. Kida$^7$}
\author{M. Hagiwara$^7$}
\author{K. Kuroki$^1$}
\author{M. Kondo$^1$}
\author{K. Tsuruda$^1$}
\author{H. Murakawa$^{1}$}
\author{N. Hanasaki$^{1}$}
\affiliation{$^1$Department of Physics, Osaka University, Toyonaka, Osaka 560-0043, Japan\\
$^2$PRESTO, Japan Science and Technology Agency, Kawaguchi, Saitama 332-0012, Japan\\
$^3$The Institute for Solid State Physics, University of Tokyo, Kashiwa, Chiba 277-8581, Japan\\
$^4$Department of Chemistry, Osaka University, Toyonaka, Osaka 560-0043, Japan\\
$^5$RIKEN Center for Emergent Matter Science (CEMS), Wako, Saitama 351-0198, Japan\\
$^6$JASRI, Sayo, Hyogo 679-5198, Japan\\
$^7$Center for Advanced High Magnetic Field Science (AHMF), Graduate School of Science, Osaka University, Toyonaka, Osaka 560-0043, Japan}
\begin{abstract}
Unconventional features of relativistic Dirac/Weyl quasi-particles in topological materials are most evidently manifested in the 2D quantum Hall effect (QHE), whose variety is further enriched by their spin and/or valley polarization.
Although its extension to three dimensions has been long-sought and inspired theoretical proposals, material candidates have been lacking.
Here we have discovered valley-contrasting spin-polarized Dirac fermions in a multilayer form in bulk antiferromagnet BaMnSb$_2$, where the out-of-plane Zeeman-type spin splitting is induced by the in-plane inversion symmetry breaking and spin-orbit coupling (SOC) in the distorted Sb square net.
Furthermore, we have observed well-defined quantized Hall plateaus together with vanishing interlayer conductivity at low temperatures as a hallmark of the half-integer QHE in a bulk form. 
The Hall conductance of each layer is found to be nearly quantized to $2(N\! +\! 1/2)e^2/h$ with $N$ being the Landau index, which is consistent with two spin-polarized Dirac valleys protected by the strong spin-valley coupling.
\end{abstract}
\maketitle
%
Researches of topological materials have currently been one of the central topics of the condensed matter physics.
Their topologically non-trivial electronic structure leads to the relativistic quasiparticles, Dirac/Weyl fermions, whose most prominent feature is seen in QHE in 2D systems\cite{Ando1982RMP}, such as the relativistic QHE in graphene\cite{Novoselov2005Nature,Zhang2005Nature} and topological insulator films\cite{Yoshimi2015NatCom}.
The half-integer quantization of the Hall plateaus and the zero-energy Landau level forming at the charge neutral Dirac point were experimentally clarified, which are associated with the Berry phase of Dirac fermions and hence have no analog in conventional 2D systems.
More recently, the variety of QHE in topological materials has been further expanded by utilizing the spin and/or valley polarization in the system\cite{Chang2013Science,Li2013PRL,Pan2014PRL,Niu20156PRB,Guinea2010NatPhys}.
The extension of such unconventional QHE to three dimension should be of particular interest in terms of novel topological phenomena\cite{Burkov2011PRL} and potential application beyond 2D systems\cite{Smejkal2017PSSR}.
However, in so-called Dirac/Weyl semimetals hosting three-dimensional relativistic quasiparticles\cite{Armitage2018RMP,Yan2017AnnuRev}, the QHE usually does not occur owing to the energy dispersion along the applied field.
%
\par
\begin{figure}
\begin{center}
\includegraphics[width=\linewidth]{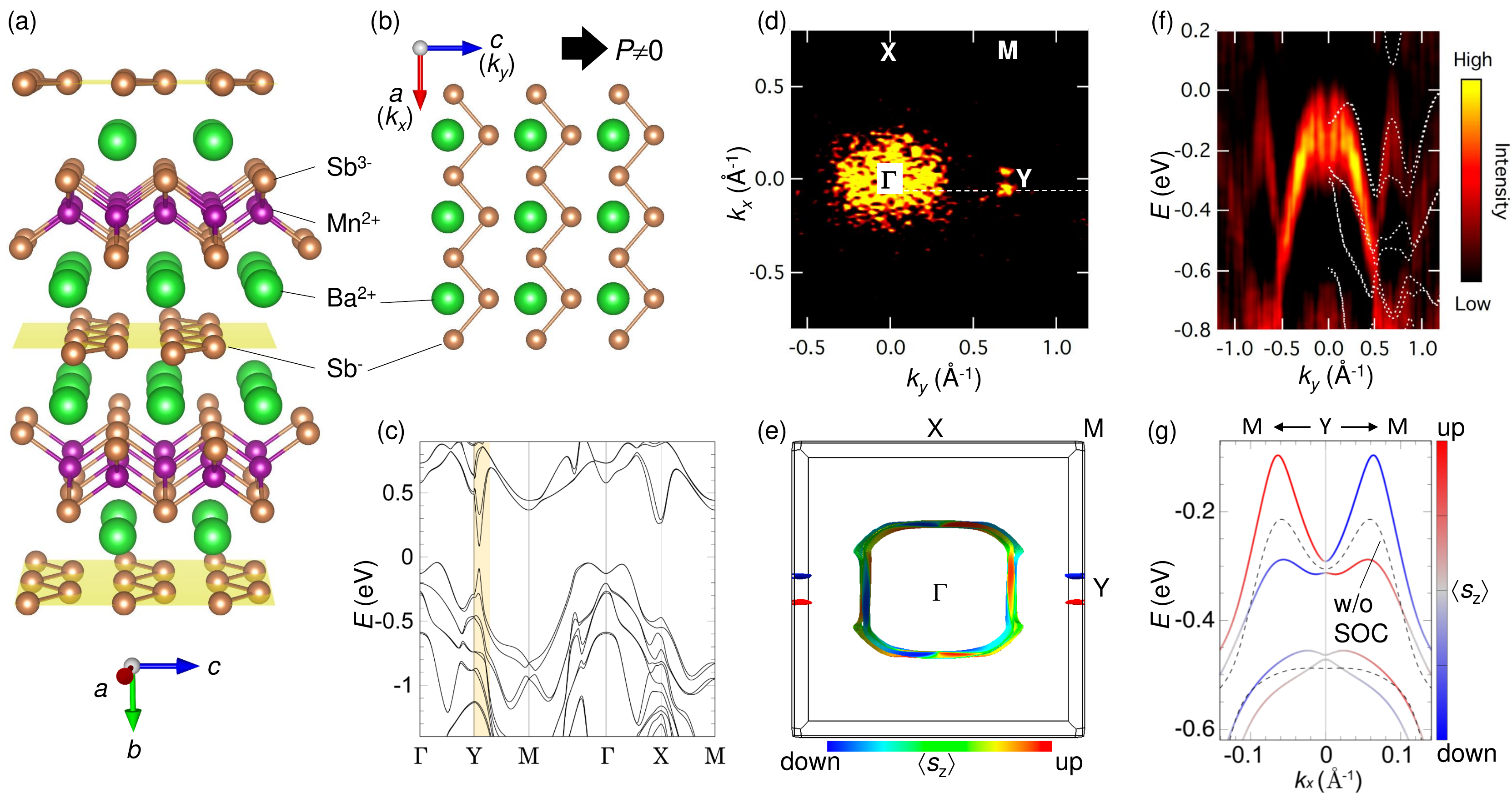}
\caption{
(Color online)
(a) Crystal structure for BaMnSb$_2$ deduced from the single-crystal x-ray diffraction at $\sim$100 K. The formal valence of each ion is also described.
(b) Plan view of the distorted Sb$^{-}$ square net (with 1D zigzag chains) coordinated with Ba ions.
(c) Calculated band structure for BaMnSb$_2$ with the SOC.
Fermi surfaces for BaMnSb$_2$, deduced from (d) the ARPES intensity map at $\sim$ 60 K (integrated over 100 meV at the Fermi level) and (e) first-principles calculation. The $k$-component along the crystallographic $a$($c$) axis corresponds to $k_x$($k_y$) [see panel (b)]. In (e), the Fermi energy is set at $-130$ meV measured from the valence band top so that the valley size matches the result of the quantum oscillation\cite{SM}.
The color of the Fermi surfaces represents the spin polarization $\langle s_z \rangle$.
(f) Cut of the ARPES intensity taken along the white dashed line in (d), with the calculation results overlaid on the right side (white dotted curve). The ARPES data are symmetrized with respect to $k_y\! =\! 0$.
(g) Calculated Dirac-like band dispersion near the Y point along the M-Y-M line, where the $\langle s_z \rangle$ value of each band is represented by red (up) and blue (down) colors. The dashed curve denotes the band dispersion calculated without the SOC.
}
\label{fig:FS}
\end{center}
\end{figure}

%
To achieve the quantum Hall states in three dimensions, multilayer systems are viable, since the interlayer coupling is so weak that the energy gap between the Landau levels can form.
In fact, over the last few decades, not only engineered semiconductor superlattices\cite{Stormer1986PRL,Druis1998PRL,Kuraguchi2000PhysicaE} but also highly-anisotropic layered compounds\cite{Cooper1989PRL,Hannahs1989PRL,Hill1998PRB,Cao2012PRL,Tajima2013PRB,Tang2019Nature} have been reported to exhibit the quantum Hall signatures.
As a topological analog, $A$Mn$X_2$ ($A$: alkaline-earth and rare-earth ions, $X$: Bi and Sb)\cite{Park2011PRL,Wang2011PRB,WangPetrovic2011PRB,JiaPRB2014,Li2016PRB,Wang2016CPB,May_EMB,Masuda_EMB,Borisenko2019NatCom,Kealhofer2018PRB,Wang2016PRB,Liu2015SR,Huang2016PNAS} is promising, since the crystal structure consists of an alternate stack of a 2D Dirac fermion conduction layer ($X^{-}$ square net)\cite{Lee2013PRB,Farhan2014JPC} and a (Mott) insulating layer ($A^{2+}$-Mn$^{2+}$-$X^{3-}$) [see Fig. \ref{fig:FS}(a)].
In fact, in addition to the $\pi$ phase shift in the SdH oscillation reported for several related materials\cite{Park2011PRL,Kealhofer2018PRB}, signature of half-integer QHE was experimentally observed in EuMnBi$_2$\cite{Masuda_EMB}.
More importantly, the $A$Mn$X_2$ materials have an advantage of enabling the QHE to be coupled with various degrees of freedom in the insulating layer, such as magnetism and lattice distortion, whicn may achieve the spin and/or valley polarization.
For instance, in EuMnBi$_2$, the exchange interaction with local Eu spins indeed causes significant spin splitting in the Dirac-like bands\cite{Masuda2018PRB}, although the external fields are necessary to induce the net magnetic moment of the Eu layers.
As another source of spin polarization, the SOC of Bi and Sb atoms is available even at zero field, if the inversion symmetry of the lattice is broken.
So far however, the crystal structure for this series of materials has been limited to a simple tetragonal one, resulting in spin degenerate Dirac bands.
%
\par
%
%
We have here discovered the in-plane inversion symmetry breaking in the weakly-distorted Sb square net of BaMnSb$_2$\cite{Liu2015SR,Huang2016PNAS,Farhan2014JPC} [Figs. \ref{fig:FS}(a) and (b)].
In combination with the SOC of Sb atoms, this polar structure causes valley-contrasting spin splitting in the Dirac-like bands at zero field, as clarified by both angle-resolved photoemission spectroscopy (ARPES) and first-principles calculations.
In addition, owing to the highly 2D nature attained by the high-temperature antiferromagnetic order of the Mn sublattice ($T_N\!\sim\! 285$ K) for BaMnSb$_2$ (Fig. S1), we have experimentally observed the half-integer QHE, where clear quantized Hall plateaus coinciding with vanishing interlayer conductivity manifest themselves at high fields even in a bulk sample.
Based on the experimental values of the quantized Hall resistivity, we estimate the spin-valley degeneracy factor consistent with the spin-valley-locking in the present compound.
%
\par
%
As a result of single-crystal x-ray structural analysis (Table S1), we have found that BaMnSb$_2$, which was previously reported to be tetragonal\cite{Liu2015SR,Huang2016PNAS}, has an orthorhombic structure, where the Sb square net is distorted to a zigzag chain-like structure [Fig. \ref{fig:FS}(b)].
The similar distorted Sb square net is formed in (Ca,Sr)MnSb$_2$~\cite{Liu2017NatMater,Ramankutty2018SciPost,You2018CAP,He2017PRB}, although the type of layer stacking is totally different; BaMnSb$_2$ hosts the coincident arrangement of Ba atoms above and below the Sb layer, while (Ca,Sr)MnSb$_2$ hosts the staggered arrangement.
Consequently, the former results in a new crystal structure (space group $Imm2$) among the $A$Mn$X_2$ family, which is polar along the in-plane direction [thick arrow along the $c$ axis in Fig. \ref{fig:FS}(b)], whereas the latter has a nonpolar space group $Pnma$.
%
\par
%
The coincident-type layer stacking for BaMnSb$_2$ is identical to that for (Sr,Eu)MnBi$_2$, which hosts quasi-2D Dirac fermions on the Bi square net.
In BaMnSb$_2$, however, the square net is distorted to form polar zigzag chains, which should greatly influence electronic states.
For clarifying this, we show in Fig. \ref{fig:FS}(c) the band structure calculated on the basis of the experimental crystal structure.
For tetragonal SrMnBi$_2$~\cite{Park2011PRL,Lee2013PRB} and EuMnBi$_2$~\cite{Borisenko2019NatCom}, the Dirac-like band along the $\Gamma$-M line crosses the Fermi energy, forming four equivalent valleys.
However, the zigzag-type distortion for orthorhombic BaMnSb$_2$ opens a sizable gap ($\sim 1$ eV) at the Dirac points on the $\Gamma$-M lines.
In the square-net layered compound, another Dirac-like band can form at the X point, where the SOC-induced energy gap tends to be large for (Sr,Eu)MnBi$_2$~\cite{Lee2013PRB} and hence no Fermi pockets are created.
For BaMnSb$_2$, on the other hand, although a similar large gap is formed at the X point, the energy gap around the Y point remains relatively small ($\sim$200 meV) even after taking account of the SOC and the zigzag-type distortion.
This leads to a highly-dispersive gapped Dirac band, as denoted by shaded area in Fig. \ref{fig:FS}(c).
The ARPES experiments have revealed that its valence band indeed crosses the Fermi level, forming two small hole pockets near the Y point along each Y-M line [Fig. \ref{fig:FS}(d)].
Note here that the broad intensity centered around the $\Gamma$ point arises from the less-dispersive parabolic valence band situated just below the Fermi level\cite{note_ARPES}.
This band is unlikely to form a Fermi surface, as is also supported by the transport features ($vide$ $infra$); the field-linear Hall resistivity and the single-frequency quantum oscillation indicate a single-carrier system.
%
\par
%
Figure \ref{fig:FS}(f) shows a cut taken along the white dashed line in Fig. \ref{fig:FS}(d) ($k_x\! =\! -0.075$\AA), where the linearly dispersing high-velocity Dirac-like band is observed near the Y point ($k_y\!\sim\! 0.7$\AA).
The ARPES dispersion is roughly reproduced by the present calculation [white dotted curves on the right side of Fig. \ref{fig:FS}(f)].
However, the calculation overestimates the energy of the parabolic band near the $\Gamma$ point relative to that of the Dirac-like band (by $\sim$100 meV).
%
\par
%
An impact of the polar lattice distortion manifests itself as valley-contrasting spin polarization of the Dirac valleys near the Y point, as shown in Figs. \ref{fig:FS}(e) and (g).
Figure \ref{fig:FS}(e) presents the calculation result of spin-resolved Fermi surfaces for BaMnSb$_2$, where two small elliptic cylindrical Fermi surfaces along the M-Y-M line are nicely reproduced.
It is striking that each Fermi surface exhibits almost full spin polarization $\langle s_z\rangle$ opposite to one another, as denoted by red (spin-up) and blue (spin-down) colors.
Note here that a large Fermi surface is formed around the $\Gamma$ point because of the overestimation of the parabolic band energy in the calculation, as discussed above.
To clarify the origin of such spin polarization depending on the valley position, we show in Fig. \ref{fig:FS}(g) the calculated band structure together with the $\langle s_z\rangle$ value along the M-Y-M line.
For comparison, the band structure without the SOC is also plotted as a dashed curve, which is less dispersive and spin-degenerate.
The onset of the SOC induces significant spin splitting in the Dirac-like band, which corresponds to the out-of-plane Zeeman-type splitting reflecting the polar structure within the plane\cite{Yuan2013NatPhys}.
Since the energy splitting is particularly large ($\sim$200 meV) around the Dirac point, only the higher-energy band hosting steep linear dispersion crosses the Fermi energy.
The $k_x$-dependent spin splitting is qualitatively explained by the SOC Hamiltonian as follows: 
$H_{\rm SO}\propto\bm{\sigma}\cdot(\bm{k}\times\bm{E})\propto \sigma_zk_xE_y$ (for $k_z\! =\! 0$), where $\bm{\sigma}$ is the spin Pauli matrix and $\bm{E}\! =\! (0, E_y,0)$ is the built-in electric field due to the polar structure.
Thus, in BaMnSb$_2$, the strong SOC together with the broken inversion symmetry inherent to the distorted Sb square net provides the spin-valley-coupled quasi-2D Dirac fermion, which is highlighted in the relativistic QHE in three dimensions shown below.
%
\par
%
\begin{figure}
\begin{center}
\includegraphics[width=.7\linewidth]{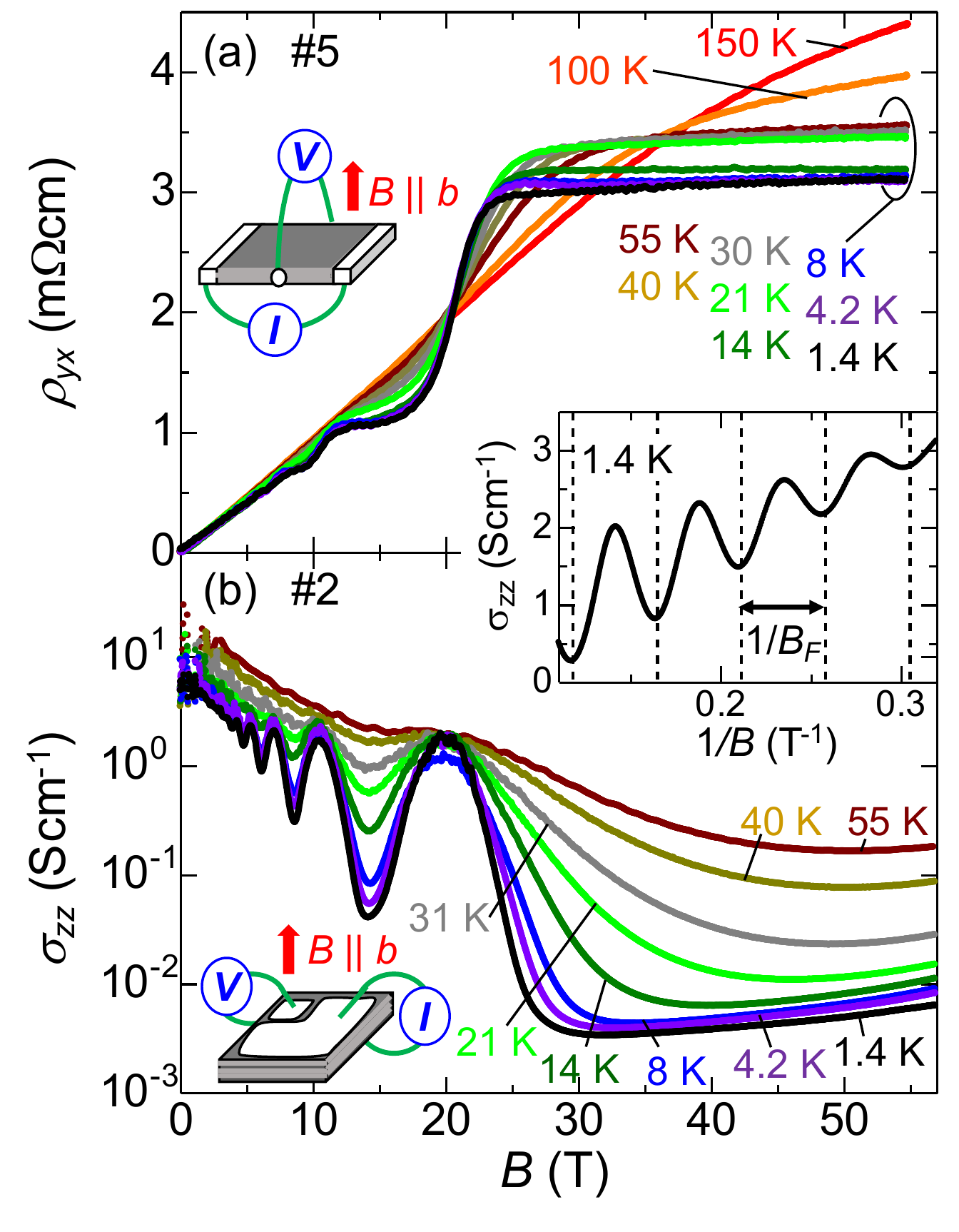}
\caption{
(Color online)
Field ($B$) dependence of (a) Hall resistivity $\rho_{yx}$ and (b) interlayer conductivity $\sigma_{zz}$ ($=\!1/\rho_{zz}$) for BaMnSb$_2$ at selected temperatures for $B||b$ up to $\sim$55 T. The measurement setup is schematically shown in each panel. Inset shows $\sigma_{zz}$ versus $1/B$ at low fields ($<9$ T), where the SdH frquency $B_F$ is defined [$B_F\!=\! 21.3(2)$ T for the sample \#2]\cite{SM}.
}
\label{fig:field}
\end{center}
\end{figure}
%
\par
%
Figure \ref{fig:field}(a) presents the field ($B$) dependence of Hall resistivity $\rho_{yx}$ for BaMnSb$_2$ at various temperatures for $B||b$.
While $\rho_{yx}$ is almost $B$ linear up to 55 T above 150 K, it weakly bends downward above 40 T at 100 K, which evolves into a clear plateau structure below 55 K.
With further decreasing temperature, in addition to the main $\rho_{yx}$ plateau above 25 T, other $\rho_{yx}$ plateaus become discernible even at low fields.
Correspondingly, we observed clear Shubnikov-de-Haas (SdH) oscillations in the interlayer conductivity $\sigma_{zz}=1/\rho_{zz}$ [Fig. \ref{fig:field}(b)], which is proportional to the density of states of the Landau levels in multilayer systems\cite{Druis1998PRL,Kuraguchi2000PhysicaE}.
To clarify the Landau levels for the present compound, $\sigma_{zz}$ is more suitable than $\sigma_{xx}$, because tiny mixture of voltage drop along the interlayer direction into that along the in-plane direction prevents the precise measurement of $\rho_{xx}$.
As temperature decreases, the dip structure of $\sigma_{zz}$ becomes deeper, leading to a giant SdH oscillation; the ratio of the peak $\sigma_{zz}$ value (at $\sim$20 T) to the bottom one (at $\sim$30 T) reaches as high as $\sim$600 at 1.4 K.
The observed vanishing $\sigma_{zz}$ accompanied by the $\rho_{yx}$ plateaus indicates that the energy gap between the Landau levels is sufficiently larger than the bandwidth of the extended state, which is the hallmark of the QHE.
%
\par
\begin{figure}
\begin{center}
\includegraphics[width=.75\linewidth]{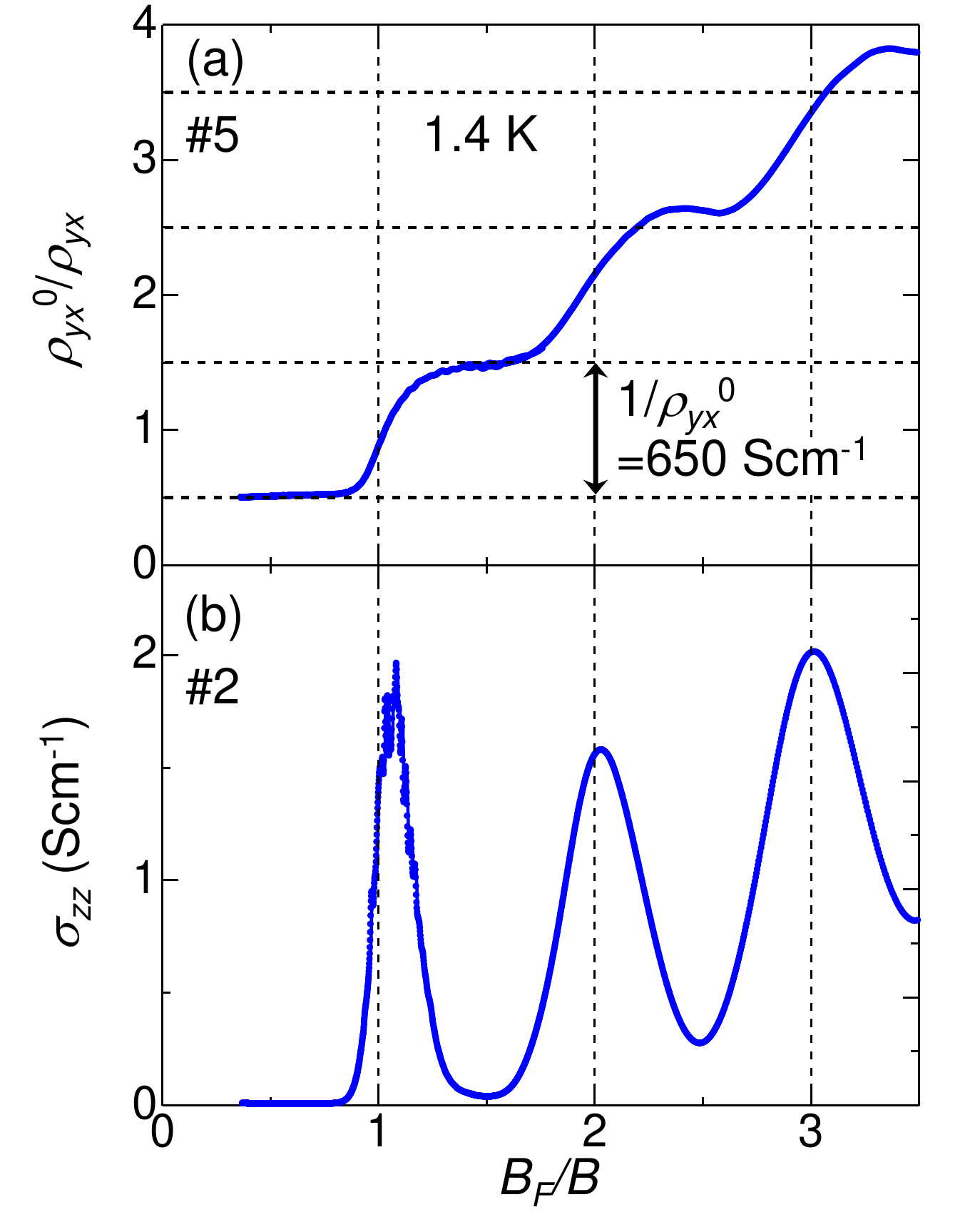}
\caption{
(Color online)
(a) Normalized inverse Hall resistivity ($\rho_{yx}^0/\rho_{yx}$) and (b) $\sigma_{zz}$ versus $B_F/B$ at 1.4 K for $B||b$, where $B_F$ is the SdH frequency estimated separately for each sample\cite{SM}. $1/\rho_{yx}^0$(=650 Scm$^{-1}$) is obtained from the step size of the $1/\rho_{yx}$ plateaus between $B_F/B\! =\! 0.5$ and $B_F/B\! =\! 1.5$, where $\sigma_{zz}$ is almost vanishing.
}
\label{fig:qhe}
\end{center}
\end{figure}

%
To show the detailed features of the QHE for BaMnSb$_2$, we plot $1/\rho_{yx}$ and $\sigma_{zz}$ at 1.4 K as a function of $B_F/B$ in Figs. \ref{fig:qhe}(a) and (b), respectively, where $B_F$ is the frequency of the SdH oscillation estimated in the low-field region [inset to Fig. \ref{fig:field}].
Note here that $B_F/B$ is the filling factor normalized by the spin-valley degeneracy factor\cite{Lukyanchuk2006PRL,Masuda_EMB}.
Clear plateau structures of $1/\rho_{yx}$ manifest themselves near half-integer values of $B_F/B$, accompanied by deep $\sigma_{zz}$ minima.
In particular, the $\sigma_{zz}$ value reaches almost zero for $B_F/B\!=\! 0.5$ and $1.5$, reflecting the well-defined quantum Hall states at high fields.
Since the quantum Hall states correspond to $B_F/B\! =\! N\! +\! 1/2\! -\!\gamma$, where $N$ is the Landau index and $\gamma$ is the phase factor expressed as $\gamma\! =\! 1/2\! -\! \phi_B/2\pi$ with $\phi_B$ the Berry phase\cite{Mikitik1999PRL}, the half-integer values of $B_F/B$ indicate $\gamma\! =\! 0$, i.e., the nontrivial $\pi$ Berry phase in the present compound.
%
\par
%
We further analyze the quantized values of $1/\rho_{yx}$.
Considering the parallel transport of each Sb layer stacking along the $b$ axis, $\rho_{yx}$ is expressed by the Hall resistance of each layer ($R_{yx}$) as follows: $\rho_{yx}\! =\! R_{yx}/Z^\ast$, where $Z^\ast\! =\! 1/(b/2)\! =\! 8.23\times 10^6$ cm$^{-1}$ is the number of the Sb layers per unit thickness with $b$ being the $b$ axis length (Table S1)\cite{Huang2016PNAS}.
For the quantum Hall state, $R_{yx}$ is quantized as $1/R_{yx}\! =\!\pm g_d\left(N\! +\! 1/2\! -\!\gamma\right)e^2/h$~\cite{Zheng2002PRB,Gusynin2005PRL}, where $e$ is the electronic charge, $h$ is the Planck's constant, $g_d$ is the spin-valley degeneracy factor.
From these relations, the $1/\rho_{yx}$ plateaus are given by $1/\rho_{yx}\!=\! Z^\ast g_d(N\! +\! 1/2\! -\!\gamma)e^2/h$, where $Z^\ast g_de^2/h\!\equiv\! 1/\rho_{yx}^0$ corresponds to the step size of successive plateaus.
Figure \ref{fig:qhe}(a) displays $1/\rho_{yx}$ scaled by $1/\rho_{yx}^0$, where the plateaus are nicely quantized to half-integer values, i.e., $N\! +\! 1/2\! -\!\gamma\!=\! 0.5$, 1.5, 2.5,...\cite{note}.
This clearly supports that each Sb layer exhibits the half-integer QHE, originating from $\phi_B\! =\! \pi$ for the Dirac fermion.
Furthermore, from the experimental value of $1/\rho_{yx}^0\! =\! 650\!\pm\! 60$ Scm$^{-1}$, we estimate the degeneracy factor $g_d$ to be $2.0\!\pm\! 0.2$, where the error includes that in the $\rho_{yx}$ (or sample thickness) measurement ($\pm$6\%, Fig. S2) and the definition of plateau positions ($\pm$3\%, Fig. S3)\cite{SM}.
The obtained $g_d$ value is thus consistent with the two spin-polarized Dirac valleys described in Figs. \ref{fig:FS}(d)~-~\ref{fig:FS}(g).
%
%
\par
\begin{figure}
\begin{center}
\includegraphics[width=\linewidth]{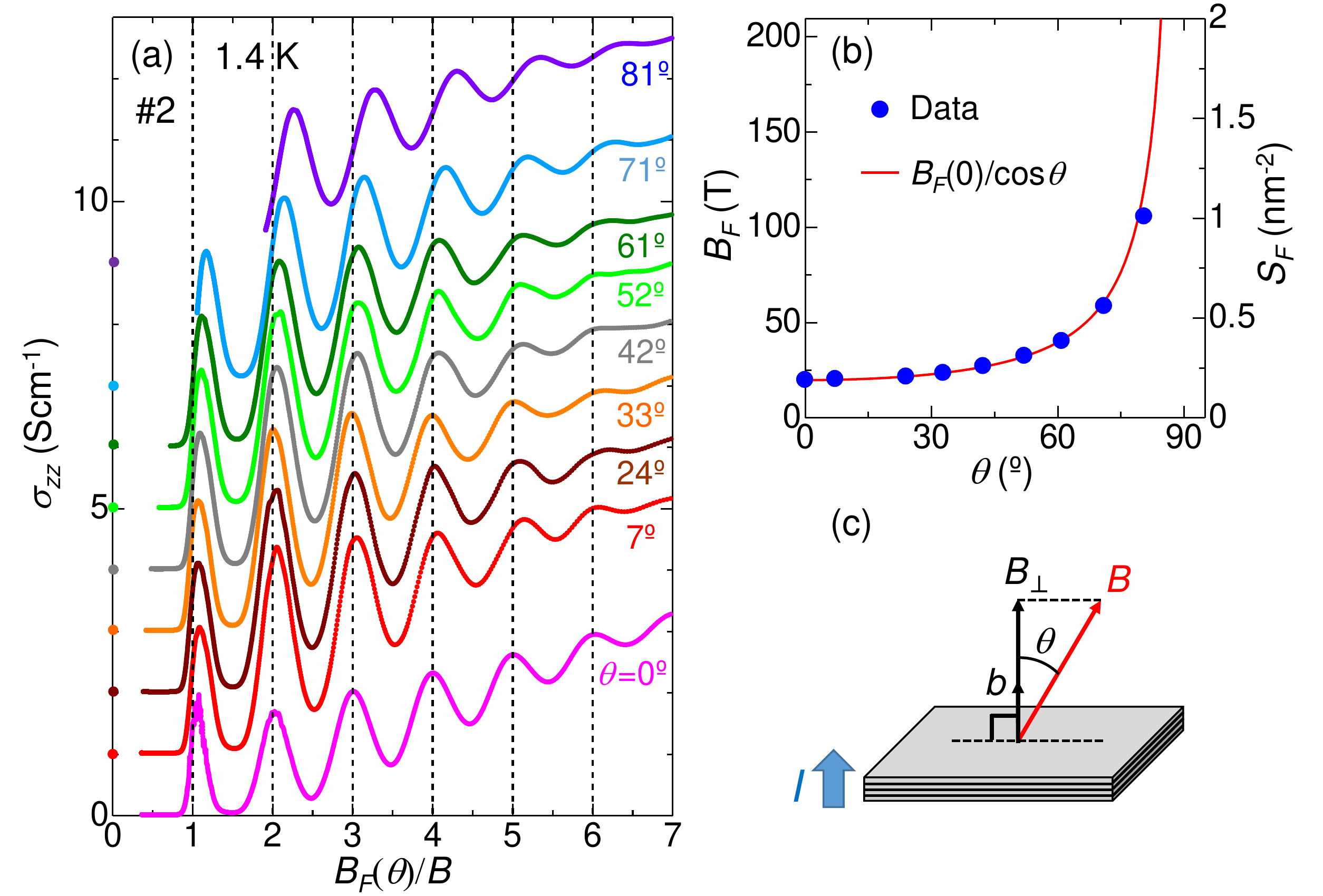}
\caption{
(Color online)
(a) Angular ($\theta$) dependence of $\sigma_{zz}$ versus $B_F(\theta)/B$ at 1.4 K, where $B_F(\theta)$ is the SdH frequency at each $\theta$. $\theta$ is the angle between the field and the $b$ axis shown in (c). Each curve at $\theta\!\ge 7^\circ$ is shifted vertically (by 1 or 2 Scm$^{-1}$) for clarity. Each origin is denoted by a closed circle on the left axis.
(b) Angular dependence of $B_F$ and the corresponding cross-sectional area of the Fermi surface $S_F$. The solid curve expresses $B_F(0)/\cos{\theta}$.
(c) Geometry of the interlayer resistivity measurement in tilted fields.
}
\label{fig:angle}
\end{center}
\end{figure}

%
To experimentally reveal the spin polarization of Dirac fermions, we measured the field angle dependence of SdH oscillations for BaMnSb$_2$.
In spin-degenerate 2D systems, in general, the ratio of the Zeeman energy $E_Z\! =\! g^\ast\mu_B B$ to the cyclotron energy $E_c\! =\! e\hbar B_\perp/m_c$ [$B_\perp\! =\! B\cos\theta$ the field component perpendicular to the 2D plane and $\theta$ the tilt angle of field from the normal to the 2D plane, see Fig. \ref{fig:angle}(c)] is tunable with $\theta$, as is given by $E_Z/E_c \!=\! g^\ast (m_c/2m_0)(1/\cos\theta)$, where $g^\ast$ is the effective $g$ factor, $\mu_B\! =\! e\hbar/2m_0$ is the Bohr magneton, $m_c$ is the cyclotron mass\cite{note3}, and $m_0$ is the bare electron mass.
Therefore, when $E_Z$ increases close to $E_c/2$ by increasing $\theta$, marked changes in amplitude and phase of the SdH oscillation are anticipated as a result of the overlap of the neighboring Landau levels with opposite spins\cite{Fang1968PR,Masuda2018PRB}.
In Fig. \ref{fig:angle}(a), we present the $\theta$ dependence of $\sigma_{zz}$ versus $B_F(\theta)/B$ at the lowest temperature, where $B_F(\theta)$ is the SdH frequency determined at each $\theta$.
At $\theta\!=\! 0^\circ$, we have $B_F(0)\!=\! 21.3(2)$ T, corresponding to the cross-sectional area of the Fermi surface normal to the field $S_F\!=\! 0.203(2)$ nm$^{-2}$, as small as $\sim$0.1\% of the total area of the Brillouin zone.
Reflecting the strong 2D nature of the Fermi surface, $B_F(\theta)$ increases with increasing $\theta$ (Fig. S4), which nicely follows the $1/\cos\theta$ scaling up to $\theta$=81$^\circ$ [Fig. \ref{fig:angle}(b)].
What is noteworthy is that the overall behavior of the SdH oscillation plotted versus $B_F(\theta)/B$ is almost independent of $\theta$, except for a slight phase shift above 71$^\circ$\cite{Tang2019Nature}.
We thus observed no spin splitting in experiments, consistent with the fully spin-polarized Dirac valleys.
%
\par
%
We note that, since two Dirac valleys are oppositely spin-polarized, the remaining valley degeneracy should be lifted by $E_Z$ when the field is applied along the $s_z$ axis ($\theta$=0$^\circ$).
As shown in Fig. \ref{fig:angle}(a), however, no signature of split Landau levels was detected at $\theta$=0$^\circ$.
This indicates $E_Z$ is much smaller than $E_c/2$, putting a constraint on the $g$ factor; $g^\ast\!\ll\! (m_0/m_c)\!\sim \! 10$, where $m_c/m_0\!\sim\! 0.1$ is deduced from the temperature dependence of the SdH oscillation at $\theta\!=\! 0^\circ$ (Fig. S5).
Such a small $g^\ast$ value for BaMnSb$_2$ may reflect the fact that the SOC-induced energy splitting dominates over $E_Z$.
If this is the case, the spins for the Dirac valleys are considered to be almost locked along the $s_z$ axis even when the field is tilted.
Therefore, $E_Z/E_c$ should be nearly independent of $\theta$ owing to $E_Z\!\propto\! B_\perp$, which is in fact consistent with the experimental results, i.e., no $\theta$-evolution of Landau level splitting up to $81^\circ$.
%
%
\par
%
In conclusion, we have found that polar antiferromagnet BaMnSb$_2$ hosts quasi-2D Dirac fermion with the Zeeman-type spin-valley locking, which originates from the in-plane inversion symmetry breaking and strong SOC in the distorted Sb square net.
We have also observed the well-defined Hall plateaus accompanied by vanishing conductivity at high fields as a manifestation of bulk QHE.
There, the peculiar Dirac fermion state is highlighted by the nearly half-integer quantization of Hall plateaus as well as the degeneracy factor reflecting the spin-polarized multiple valleys.
Thus, the $A$Mn$X_2$ compounds hosting a distorted $X$ square net can be a novel platform for a variety of spin-valley-coupled Dirac fermions, which open the door to exotic QHE phenomena in a bulk form that may contribute to emerging topological (spin)electronics.
%
\begin{acknowledgments}
The authors thank N. Kumada, Y. Fuseya, T. Osada, M. Koshino, Y. Niimi, H. Suzuura, and T. Arima for helpful discussions.
This work was partly supported by the JST PRESTO (Grant No. JPMJPR16R2), the JSPS KAKENHI (Grant Nos. 16H06015, 19H01851, 19K21851, JP19H05173) and the Asahi Glass Foundation.
The synchrotron radiation experiments were performed at the BL25SU of SPring-8 with the approval of the Japan Synchrotron Radiation Research Institute (JASRI) (Proposals No. 2019A1087).
This work was partly carried out at AHMF in Osaka University under the Visiting Researcher's Program of the Institute for Solid State Physics, the University of Tokyo and at the Center for Spintronics Research Network (CSRN), Graduate School of Engineering Science, Osaka University.
\end{acknowledgments}
%


\begin{thebibliography}{}
\bibitem{note_ARPES} The valence band, whose top is close to the Fermi level, can give rise to intensity, since the ARPES mapping was obtained by integrating over 100 meV.
\bibitem{note} The quantized values of $\rho_{yx}^0/\rho_{yx}$ slightly deviate from half-integer values for $B_F/B\ge 2.5$, which presumably stems from the nonzero values of the longitudinal conductivity (or resistivity) minima at low fields.
\bibitem{note3} Since $E_c$ is dependent on $N$ for a 2D Dirac fermion, we here effectively define $E_c$ by using a semiclassical expression of the cyclotron mass $m_c\! =\! E_F/v_F^2$ with $v_F$ and $E_F$ being the Fermi velocity and Fermi energy, respectively.
\end{thebibliography}
\end{document}